\documentclass[twocolumn]{revtex4}
\usepackage{graphicx}

\begin{document}

\title{Finite size effects in the dynamics of opinion formation}
\author{ Ra\'{u}l Toral and Claudio J. Tessone}

\address{Instituto Mediterr\'{a}neo de Estudios Avanzados (IMEDEA), CSIC-UIB, Ed. Mateu Orfila, Campus UIB, E-07122 Palma de Mallorca, Spain}

\begin{abstract}
For some models of relevance in the social sciences we review some examples in which system size plays an important role in the final outcome of the dynamics. We discuss the conditions under which changes of behavior can appear only when the number of agents in the model takes a finite value. Those changes of behavior can be related to the apparent phase transitions that appear in some physical models.  We show examples in the Galam's model of opinion transmission and the Axelrod's model of culture formation stressing the role that the network of interactions has on the main results of both models. Finally, we present the phenomenon of system-size stochastic resonance by which a forcing signal (identified as an advertising agent) is optimally amplified by a population of the right (intermediate) size. Our work stresses the role that the system size has in the dynamics of social systems and the inappropriateness of taking the thermodynamic limit for these systems.
\end{abstract}
\date{\today} \maketitle

\section{Introduction}
\label{introduction}
In statistical physics we are used to taking routinely the thermodynamic limit in which the number of constituents (or, more precisely, the number of degrees of freedom) $N$ tends to infinity\cite{pathria}. This is necessary in many aspects, such as determining the validity of the statistical approach, the equivalence between ensembles, the existence of phase transitions with symmetry breaking and discontinuities in the order parameters or their derivatives, etc. When we apply the methods of statistical physics to explain the properties of macroscopic matter, it is clear that the number $N$ is always finite, but very large. A good estimative of its order of magnitude is the Avogadro number $N_0=6.023\times10^{23}$. Consequently, the nowadays widely used computer simulations of physical systems always struggle to get to larger and larger systems with the continuous increasing demand in computer resources. In many cases as, for instance, when trying to predict the thermodynamical properties near a critical point, it is essential to use the well developed techniques of finite size scaling\cite{cardy} in order to extrapolate the results to $N\to\infty$. 

When applying the same tools of statistical physics to problems of interest in social sciences \cite{weidlich,Ball,stauffer04}, we have to take into account that the number of individuals or agents considered can never be that large. In most cases, realistic values of $N$ range in the hundreds or thousands, reaching at most a few million. The thermodynamic limit might no be justified in this case, as the results in that limit can vary with respect to those of finite-size systems. Furthermore, new and interesting phenomena can appear depending on the number of individuals or agents considered. In this paper we will review some finite-size related effects that have been recently found in some models of interest in social systems and that exemplify cases in which the thermodynamic limit does not capture the essentials of the system's behavior. 

The outline of the paper is as follows: in the following section \ref{sec:apparent}, we will explain the concept of {\sl apparent phase transitions} and {\sl pseudo-critical points} by reviewing a physical example in which a clearly observable change of behavior disappears in the thermodynamic limit. We will then show in the following sections that apparent phase transitions also appear in two widely used models: the Galam model for opinion spreading (in section \ref{sec:galam}) and the Axelrod model of culture formation (in section \ref{sec:axelrod}); finally, section \ref{sec:sssr}  is devoted to the phenomenon of system-size stochastic resonance in a model for opinion formation that takes into account external influences. A brief summary and concluding remarks are presented in the final section \ref{sec:conclusions}.

\section{Apparent phase transitions}

\subsection{Pseudo-critical temperatures}
\label{sec:apparent}
A true phase transition can only appear in the thermodynamic limit \cite{pathria}. This limit is needed for the singularities inherent to the transition (for instance, the divergences in the critical point of a continuous, second order, phase transition or the discontinuities in a first order phase transition) to develop from the sum of analytic functions. However, in some cases, finite systems show a behavior that reminds us of a phase transition. An example which has been extensively studied  is the sine--Gordon model for the growth of surfaces \cite{schneider,TS}. Nowadays, it is well established that in two or more spatial dimensions the model displays a Kosterlitz--Thouless-type phase transition from a (low temperature) smooth phase to a (high temperature) rough phase at a given critical value $T_c$ of the temperature \cite{Falo,sanbimo,CW2,knopf,nozi}. 

\begin{figure}[t!]
\begin{center}
 \includegraphics[width=8.5cm]{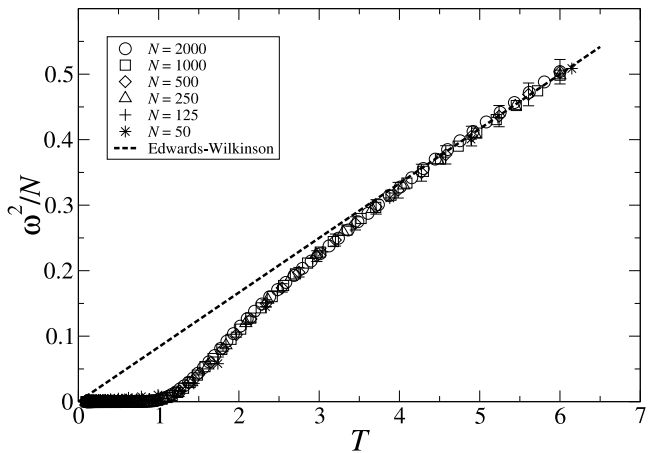}
\caption{\label{fig:rough}Roughness $\omega^2/N$ of the one-dimensional sine-Gordon model as a function of the temperature \cite{acst03}. Observe the apparent transition between a smooth phase (roughness equal to zero) to a rough phase at a pseudo-critical temperature $T_c\approx 1$. Different symbols correspond to different system sizes, as indicated in the figure and the line is the result $\omega^2/N=T/12$ of the Edwards-Wilkinson model valid at high temperatures \cite{ft93}. }
\end{center}
\end{figure}

However, numerical simulations of this model seemed to imply that the same transition was also present in the one-dimensional case \cite{acst03}. See the evidence in figure \ref{fig:rough} where it can be observed that the roughness of the one-dimensional surface is zero until a temperature $T_c \approx 1$ and then increases suddenly. Despite this numerical evidence, the result was doubtful, since the theory clearly indicated that such a phase transition was not possible for a one-dimensional system \cite{vanHove,lieb,CS}. In the view of this negative prediction, one can revisit the simulation data and conclude \cite{acst03} that the deviation from smooth to rough surfaces occurs at a pseudo-critical temperature that depends on the system size as $T_c(N)\sim 1/\ln(N)$. It is certainly true that in the thermodynamic limit the transition point switches to $T_c=0$. In other words, the system does not display a phase transition, as demanded by the rigourous theorems. However, it is also true that for every finite system, a well defined transition point (which is not a true critical one) can be identified as separating the smooth from the rough behavior. Since the dynamics at low temperature is dominated by the existence of many metastable states with large energy barriers \cite{schneider}, a lot of numerical effort is needed in order to proof beyond any doubts that the transition point tends to zero for increasing system size\cite{acst03}.

\subsection{Galam's model}
\label{sec:galam}
There are several models that analyze the dynamics of opinion formation in a society \cite{Gal03,Gal00,SG:2002,SG:2003,complex1,complex1b,heg01,sznajd01,sznajd01b,sznajd01c,sznajd02,stauffer1,stauffer1b,stauffer11,stauffer2,stauffer2b,stauffer2c,Castellano,castellano2,Klemm1,Klemm2,Klemm3,krapi01,Mobilia1}. The model introduced by S. Galam \cite{SG:2002,SG:2003} aims to identify mechanisms by which an opinion, being initially minority, can become majority in the long run. There have been many recent examples of such change of opinion: the rejection to the Maastricht treaty in Ireland, the negative vote in France for the European constitution, the switch in the belief of the authorship of the March 11th 2004 terrorists attacks in Madrid, the rumors concerning some September 11th opinions in France \cite{SG:2003}, etc. The basic ingredient of the model is that the opinion a social group has on a topic changes when the individuals meet to discuss in small groups. To simplify matters, the model considers that the opinion of an individual can only take two possible values: in favor or against. After discussion, the majority opinion is adopted by all the members of the discussion group. In case of a tie, i.e.~same number of individuals against and in favor, one of the two possible decisions is always favored and hence adopted. This favored decision usually responds to some sort of social inertia against change. For instance, when discussing about the reforms in the European constitution, the opinion favored by the French people was that of voting against its approval because it was unclear the advantages that the new constitution would bring to the French society. The model has been recently modified in order to acount for the diffusion of the individuals \cite{stauffer11} and the effects of {\sl contrarians} (people holding a different opinion of that of the majority) \cite{Gal04b,stauffer4}.

\begin{figure}[t!]
\begin{center}
 \includegraphics[width=6.5cm,angle=-90]{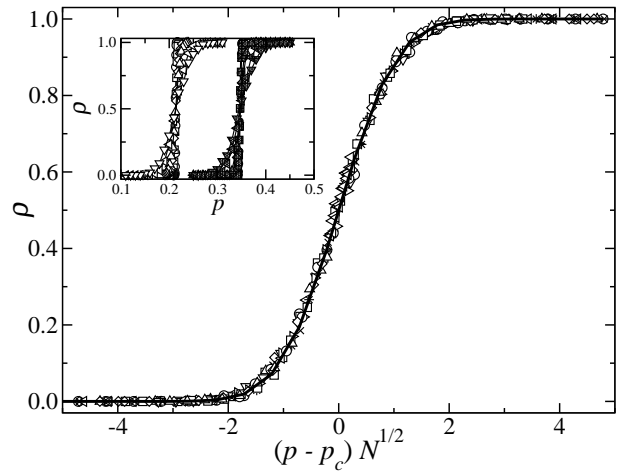}
\caption{\label{rhopn}The inset shows $\rho$ as a function of $p$ and $N$ for Galam's model in which groups form independently of any newtork structure. The size distribution of discussion groups is uniformly distributed in $[1,M]$. Open symbols correspond to $M=4$ ($p_c=0.2133$), while filled ones to $M=5$ ($p_c=0.3468$). Different symbols correspond to different system sizes between
$N=10^3$ and $N=10^6$ (the transition sharpens with increasing system size). The main plot shows the collapse according to the scaling function $\rho(p,N) = (1+\hbox{erf}(x/1.17))/2 $ as a function of the scaling variable $x=(p-p_c)N^{1/2}$. }
\end{center}
\end{figure}

If one assumes that the discussion groups are formed by taking at random individuals from the whole society, it is easy to establish a recurrence relation for the fraction of people $P_+(t)$ holding the favored opinion at time $t$. This assumption by which any two individuals have the same chance to meet in a discussion group belongs to the category of {\sl mean-field} approaches, widely used as a first approximation in many problems of interacting particles\cite{stanley}. The recurrence relation adopts the general form: $P_+(t+1)=F[P_+(t)]$, where $F$ is some polynomial function that depends on the size distribution of the discussion groups \cite{SG:2003}. The analysis of this recurrence relation shows that there are three fixed points: the trivial ones (which are stable) $P_+=0$ and $P_+=1$, and a non-trivial {\sl faith} point $P_+=p_c$ (an unstable fixed point). Let us denote by $p$ the initial proportion of supporters $p=P_+(0)$ of the favored opinion. If this proportion is greater than the faith-point value, $p>p_c$, then the recurrence relation tends to the fixed point $P_+=1$ and the whole population adopts the favored opinion. Otherwise, if $P_+(0)<p_c$ the fraction of people supporting the favored opinion tends to $0$ as the recurrence relation proceeds in time.
If we define an order parameter $\rho = \langle P_+(t\to\infty) \rangle$ as the average, over realizations of the discussion process, of the fraction of people holding the favored opinion after many discussion steps, the predictions of this simple analysis is that
\begin{equation}
\rho(p)=\left\{ \begin{array}{ll}
1 & {\rm if\, } p>p_c \\ 
0 & {\rm if\, } p<p_c.
\end{array} \right.
\end{equation}
Since $p_c<1/2$ it is possible that a favored opinion that was held initially by a minority, $p_c<p<1/2$, is in the long run the opinion hold by the whole population. That would be the explanation, according to this model, of the fact that an initially minority opinion can become majority if the society is allowed to discuss about it for a long enough time.

This analysis, based on the stability of the fixed points, is only valid in the $N\to \infty$ limit. In fact, for a small population, the condition $p>p_c$ does not automatically ensure that the preferred opinion will be the favored one. There are finite size fluctuations such that the actual result of the repeated discussion process is not well established if the difference between the initial fraction of supporters and the faith point is of order $|p-p_c|\sim N^{-1/2}$. This rounding-off of the sharp transition can be summarized in the following finite-size scaling law:
\begin{equation}
\rho(p, N) = \rho \left( ( p-p_c ) \, N^{1/2} \right),
\end{equation}
a result well confirmed by the computer simulations, as shown in figure \ref{rhopn}.

\begin{figure}[t!]
\begin{center}
 \includegraphics[width=6.5cm,angle=-90]{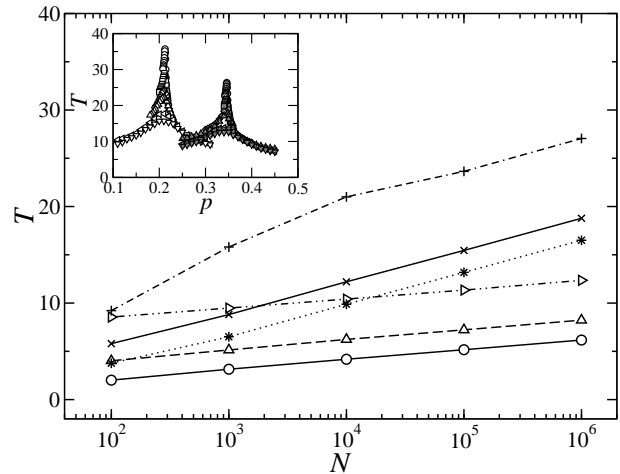}
\caption{\label{timepn} Time $T$ to reach consensus as a function of $N$ in the Galam's original model for different values of $p$ and a maximum group size of $M=5$. Respectively from top to bottom (in the right end of the graph) the different curves show $p = 0.20, 0.10, 0.05, 0.30, 0.60, 0.90$. In the inset, the time $T$ is plotted as a function of $p$ for the same system parameter as in figure \ref{rhopn}, with the same symbols meaning than in that figure. Observe the  slowing-down for $p$ close to the critical points.}
\end{center}
\end{figure}

It is possible to compute the time it takes the population to reach the final consensus. According to Galam's analysis, the recurrence relation reaches a {\sl de facto} consensus after a finite, small, number of iterations. For a finite-size population, this can be quantified precisely \cite{TTAWS:2004} as the number of steps $T$ required for $P_{+}(T) = 1/N$ or $P_+(T)=1-1/N$ indicating that the vast majority of the population has adopted the same opinion. In figure \ref{timepn} it is shown that this time scales as $T \sim \ln N$. Numerical simulations, as can be seen in the same figure, also show that the time to reach consensus grows near the faith point; this slwowing down can be understood as initially the system escapes very slowly from the unstable fixed point $p_c$. Therefore, the conclusion of this mean-field analysis is that this model displays a true phase transition between regimes dominated by one or the other of the two possible opinions. The transition is rounded-off for a finite system and finite-size scaling laws can be used to analyze the results in finite systems.
 
More dramatic finite size-effects appear if we go beyond the mean-field theory and consider spatial effects within the so-called {\sl neighborhood models} \cite{TTAWS:2004}. Contrary to the previous scheme in which an individual could join any discussion group, neighborhood models assume that discussion groups are formed with some criterion of locality or geographical neighborhood. Some other spatial effects of this model have been discussed in references  \cite{galam1,galam2}. Here we consider that the $N$ individuals are fixed at the sites of a regular lattice and the discussion cells are well defined closed, localized, rectangular regions. The exact dynamics is as follows: at each time step, an individual is chosen and a rectangle with length sides $m_x$ and $m_y$ is built around that individual. Both numbers, $m_x$ and $m_y$, are chosen uniformly from an interval $[1,M]$. The $m_x\times m_y$ individuals on this region discuss about the selected topic and all reach a local consensus based upon the majority rule, with a bias (favored opinion) in case of a tie. The opinion of all individuals in the group is changed accordingly and time increases by an amount $t\to t+m_x\times m_y/N$.

 \begin{figure}[t!]
\begin{center}
 \includegraphics[width=6.5cm,angle=-90]{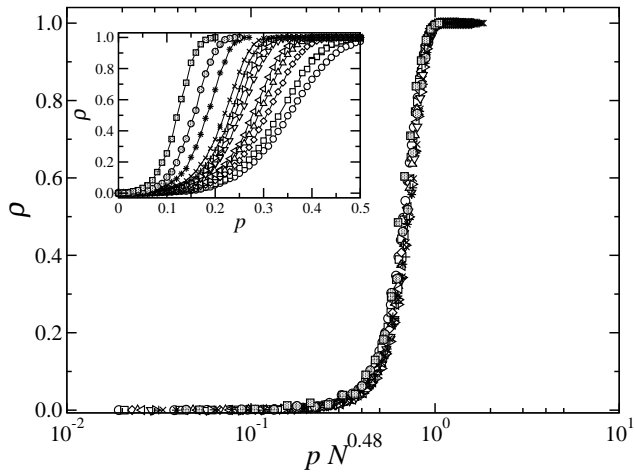}
\caption{\label{rhopn2}$\rho$ as a function of $p$ for different values of $N$ in the neighborhood's Galam model. The maximum lateral size of the groups is $M=5$. The system size varies from $N=225$ to $N=10^6$. The main plot shows the scaling law $\rho(p,N)=\rho(p\,N^{\alpha})$ with $\alpha=0.48$. The inset shows the unscaled results, showing the transition point shifting toward $p=0$ as size increases.}
\end{center}
\end{figure}
 
As in the original Galam's model, a global consensus is always reached in a finite time. The first feature that appears in the numerical simulations of this neighborhood model is the smooth transition between the two different selected states as a function of the initial proportion $p$ of supporters of the favored opinion. This is evident in the inset of figure \ref{rhopn2}, where we plot the order parameter $\rho$ as a function of $p$. This looks similar to the rounding-off induced by finite size of the original model. However, the new feature is that the transition point, defined as the fraction $p$ for which $\rho=1/2$, clearly decreases with system size $N$. In fact, a finite size scaling analysis shows that the data can be well fitted by the formula:
\begin{equation}
\rho(p,N)=\rho(p\,N^{\alpha})
\end{equation}
with an exponent $\alpha\approx 0.48$ for the dynamical rules detailed above. The transition point scales hence as $p_c(N)\sim N^{-\alpha}$. An strict statistical mechanics analysis based on the thermodynamic limit $N\to\infty$ would conclude that the {\sl actual} critical point is $p_c=0$ and hence the transition between the two regimes has disappeared! In our opinion, however, this would be a misleading conclusion. There is indeed a well defined transition between the dominance of the two consensus opinions for a finite number $N$,  of the order of magnitude to be encountered when applying this model to any situation of interest. In an infinite system, the conclusion is that the favored opinion eventually dominates regardless the (infinitesimally small) amount of initial supporters. In any finite system, though, it is possible to define a transition between the two regimes at a value $p_c(N)$. Equivalently, for a fixed proportion of initial supporters $p$ one could identify a value $N_c(p)$ separating the different consensus obtained whenever $N>N_c(p)$ or $N<N_c(p)$. Hence, the change in behavior is induced by the variation of the number of individuals, $N$. A result that would have been missed if we had taken the thermodynamic limit.

\begin{figure}[t!]
\begin{center}
 \includegraphics[width=6.5cm,angle=-90]{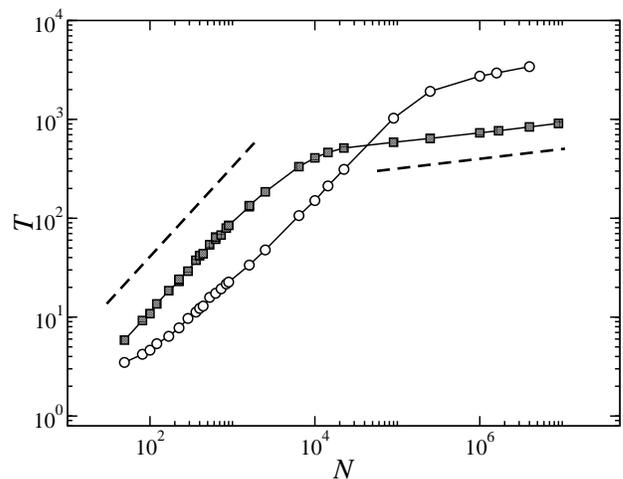}
\caption{\label{timepn2} Time to reach consensus as a function of $N$ in the neighborhood version of Galam's model, for $p=0.30$ (circles) and $p=0.20$ (squares). Two different scalig regimes for large and small $N$ can be clearly seen.}
\end{center}
\end{figure}

For the neighborhod model, it is possible also to compute the time $T$ it takes to reach the final consensus. Again, the simulations indicate that this time increases near the above defined transition point $p_c(N)$, and this {\sl critical slowing down} is a typical signature of the existence of a phase transition. In this case, see figure \ref{timepn2}, the characteristic time scales as $T(N)\sim N^{\beta}$ with $\beta\sim 0.6$ for $N<N_c(p)$ and $\beta\sim 0.1$ for $N>N_c(p)$. The reason for these two regimes to appear is that, as discussed above, the final state to which the system approaches depends on system size: for finite $p<1/2$ and small $N$ the system will approach to $P_+=0$ (as $\rho=0$), but for a large enough system, the absorbing state will be $P_+=1$, and thus the dependence of $T$ with system size experiments a strong change.

\subsection{Axelrod's model}\label{sec:axelrod}
In an important paper \cite{Axelrod97} R.~Axelrod has introduced a model for the generation and diversification of different cultures. The basic question addressed by Axelrod is the observation that differences between cultures do not disappear at all despite the fact that people tend to become more alike in their beliefs, attitudes and behaviors when they interact. His proposal to explain this paradox is a model to explore mechanisms of competition between {\sl globalization} (understood as a uniform culture) and {\sl polarization} (understood as the persistence of cultural diversity). 

In the Axelrod model, {\sl culture} is defined as a set of attributes subject to social influence. Each one of these attributes can take several traits. The traits change by social interaction between neighboring agents. The main premise is that the more similar an agent is to a neighbor, the more likely the agent will adopt one of the neighbor's traits. In this way, the model incorporates the observation that the communication is most effective between already similar people. The novelty this model brings into social modeling is that it takes into account the interaction between different cultural features. 

More specifically, Axelrod's model considers that each agent is characterized by a set of $F$ cultural features, each one of them can adopt any of $q$ different traits. The state of agent $i$ is hence characterized by the $F$ variables $(\sigma_{i1},\dots,\sigma_{iF})$, each one of them can take the values $\sigma_{ik}\in(0,\dots,q-1)$. The number of possible traits per feature $q$ can be considered as some sort of measure of the degree of heterogeneity a cultural feature might have. Agents are located in the sites of a lattice. The connectivity between the different lattice sites defines the neighbor structure of the social network. The dynamics is as follows: first, two neighbor agents $i$ and $j$ are randomly selected. One then computes their {\sl overlap} $\ell_{ij}$ or number of common features, $\ell_{ij}=\sum_k\delta_{\sigma_{ik},\sigma_{jk}}$. With a probability $\ell_{ij}/F$, the value of one of the not yet common features is transferred from one agent to the other, so increasing the overlap by one. This process is repeated by randomly selecting another couple of neighboring agents. Eventually, a frozen state is reached in which no possible evolution is possible. In such a frozen state, neighboring sites have either an overlap equal to $0$ or to $F$. The relative size of the largest cluster of agents sharing all cultural features, $S_{max}/N$ is a measure of the cultural diversity. If $S_{max}=N$ it means that all agents share the same cultural features, a monocultural or {\sl globalized} state. If $S_{max}=1$ it means that no neighboring agents share the same culture, a state that can be qualified as of cultural diversity or {\sl polarized}. 

For fixed value of $F$ it was shown by Castellano {\it et al.} \cite{Castellano,castellano2} that a genuine phase transition occurs between polarized and globalized states at a critical value, $q_c(F)$, of the parameter $q$. For $q<q_c(F)$ a monocultural state appears, whereas for $q>q_c(F)$ the society splits in several regions each one with a different culture. 

\begin{figure}[t!]
\begin{center}
 \includegraphics[width=6.5cm,angle=-90]{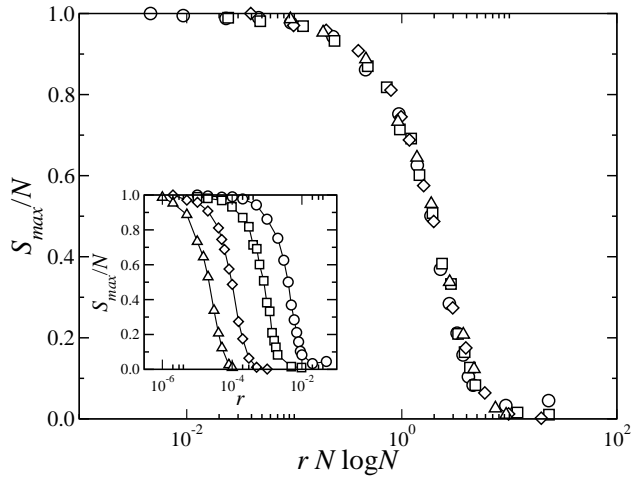}
\caption{\label{axel2} For the Axelrod model defined in a regular two-dimensional lattice with nearest neighbors interactions, we plot the relative size of the largest cluster $S_{max}/N$ as a function of the probability of a random change in a cultural trait, $r$. Other parameters are $F=10$, $q=100$. In the inset, we plot the raw results for different system sizes $N=10^2, 20^2, 50^2, 100^2$ shown with circles, squares, diamonds and triangles, respectively. The main plot shows the rescaled results, showing a data collapse when the $x$-axis is rescaled as $r \, N\ln N$.}
\end{center}
\end{figure}

\begin{figure}[t!]
\begin{center}
 \includegraphics[width=6.5cm,angle=-90]{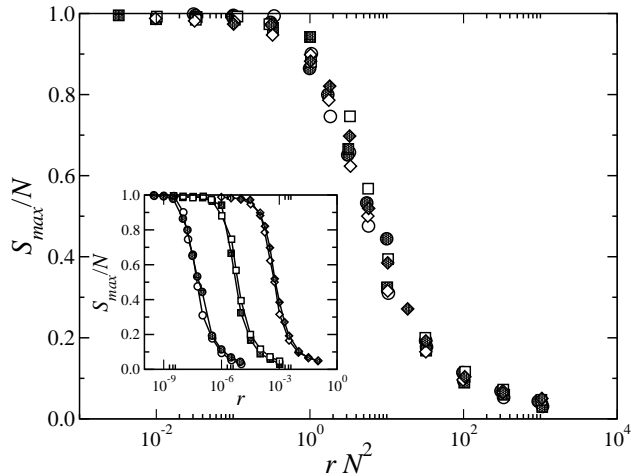}
\caption{\label{axel1} Same as figure \ref{axel2} in the case of a one-dimensional lattice. The necessary scaling factor for the noise intensity is now $r \, N^2$. The symbols correspond to different system sizes $N=10^4, 10^3, 10^2$ shown with circles, squares and diamonds, respectively. Filled symbols use $q=5$ while open symbols correspond to $q=10$. In all cases it sis $F=10$.}
\end{center}
\end{figure}

The analysis of Castellano {\it et al.} corresponds to the situation in which fluctuations are neglected. The situation is dramatically altered when one considers the effect that {\sl cultural drift} has on the dynamical evolution. Quoting Axelrod: ``Perhaps the most interesting extension and at the same time, the most difficult one to analyze is cultural drift (modeled as spontaneous change in a trait)" \cite{AxelrodBook}. We have modeled cultural drift by adding one step to the original Axelrod model: with probability $r$ an agent changes randomly one of his cultural traits. For fixed values of $F$ and $q$ a change of behavior can be clearly observed when varying the rate $r$. A {\sl pseudo-critical} value $r_c$ can be defined such that if $r>r_c$ the final state is multicultural, whereas for $r<r_c$ the cultural drift can not lead the system to develop many cultural states\footnote{Note that for $r<r_c$, the system is most of the time in a single monocultural state, but it eventually switches to another monocultural state in a really dynamical equilibrium.}. Note that we have used again the word {\sl pseudo-critical} to denote the value $r_c$. The reason, as before, is that $r_c$ depends on the system size $N$. The exact dependence is related to the dimensionality of the underlying lattice. For a regular two-dimensional network of interactions, it is $r_c\sim1/N\ln N$\cite{Klemm2}, see figure \ref{axel2}, whereas in a one-dimensional lattice the dependence is $r_c\sim 1/N^2$\cite{Klemm3} as seen in  figure \ref{axel1}. In both cases, in the limit of infinite size $N\to\infty$, the value of $r_c$ tends to $0$. In this limit, for any positive value of $r$ it is $r>r_c$ and the system tends to a multicultural state independently of any other parameters. In the language of critical phenomena, the cultural drift is a {\sl relevant} variable since any vanishingly small value alters the final state of equilibrium. However, in a finite system there is always a well defined value $r_c$ which separates the transition between polarization and globalization. Again, an analysis based upon the infinite size limit would have missed this interesting property of Axelrod's model.

\begin{figure}[t!]
\begin{center}
 \includegraphics[width=6.5cm,angle=-90]{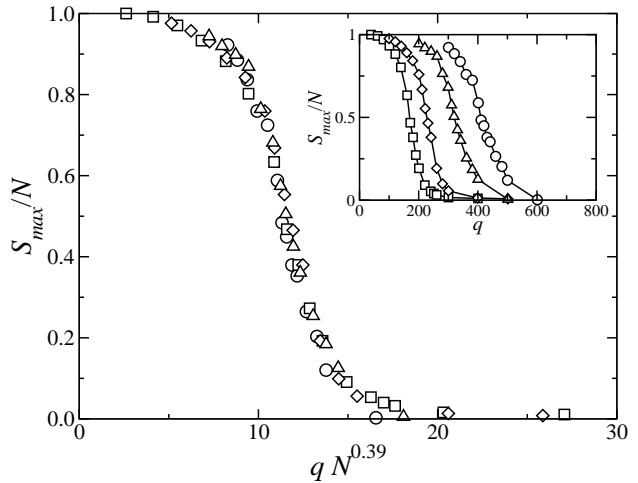}
\caption{\label{axel3} Axelrod model in a Barab\'asi-Albert lattice. The necessary scaling factor for the noise intensity is now $r \, N^{0.39}$. The symbols correspond to different system sizes $N=10^3, 2\times 10^3,, 5\times 10^3, 10^4$ shown with circles, squares, diamonds and triangles, respectively, and $F=10$.}
\end{center}
\end{figure}

Another interesting modification of the Axelrod model considers complex networks of interactions between the agents.  In order to mimic some of the conditions that a social network has, the connectivity is that of a {\sl small world} network introduced by Watts and Strogatz \cite{WS98}: individuals are considered to be located in the sites of a regular network with nearest-neighbors interaction; with a probability $p$ one of the nearest neighbor links is rewired and points to another, randomly selected individual, so establishing a long-distance connection. In this small-world network, the phase transition observed by Castellano {\sl et al.} is still present, although its exact location shifts to larger values of the heterogeneity parameter $q$ as the rewiring parameter $p$ increases. This shows that, as expected, the enhanced connectivity of the small-world network favors the globalized state. A different scenario appears in the scale-free networks. Those networks are characterized by a power law tail in the distribution of the number of neighbors. The first example is that of the Barab\'asi-Albert network\cite{Barabasi99a} which is grown using the mechanisms of preferential attachment of links. In the Barab\'asi-Albert network, the location of the (pseudo-)critical value $q_c$ depends on system size as $q_c(N)\propto N^{-\alpha}$ with $\alpha\sim 0.39$. This is a consequence of the finite-size scaling law by which the order parameter $S_{max}/N$ turns out to be a function of the product $qN^{\alpha}$, as evidenced in figure \ref{axel3}. With the thermodynamic limit in mind, the conclusion is that the transition betweeen globalized and polarized states disappears in this case. However, as discussed before, this conclusion does not prevent the finite system from having a true, observable, change of behavior as the parameter $q$ varies. It should be said that the transition is recovered, even in the thermodynamic limit, when one considers other types of more complex networks such as the structured-scale-free networks\cite{Klemm3,victornet1}.

\section{System size stochastic resonance}\label{sec:sssr}

It was found more than two decades ago that stochastic terms ({\sl noise}) can have a constructive role in non-linear dynamical systems, \cite{NN:1981,BSV:1981}. In these pioneer works, it was shown that the response of a dynamical system to a weak signal can be {\em improved} in presence of the right amount of noise, but when the strength of the stochastic terms is too large or too small, such improvement disappears. This phenomenon receives the name of {\em stochastic resonance}. 

The original motivation of these works was to provide a framework in which the small variations in the radiation reaching the Earth from the Sun could be amplified to account for the big climatic changes observed in geological ages. However, stochastic resonance \cite{JSP70,1998_Gammaitoni_reviews} is now known to hold in a large variety of systems subjected to many different kinds of forcing. The paradigmatic example is that of an stochastic dynamics in a bistable potential under the effect of a weak periodic signal. In the absence of the forcing signal, the stochastic dynamics is such that there are jumps between the two stable wells of the potential with a characteristic Kramers time, $T_K$\cite{kramers}. The role of the forcing is to lower periodically the barrier to go from one well to the other in such a way that it helps the stochastic dynamics to overcome that barrier alternatively in one direction or another following the forcing. If the value of half the period $T/2$ coincides with the Kramers time, then the external signal is optimally followed by the stochastic dynamics. The matching $T_K=T/2$ can be achieved by varying the intensity of the noise, in such a way that an optimal response to the action of the periodic forcing is obtained for the right amount of noise.
 
The very general conditions under which the phenomenon can appear has opened a very active field of research. Many other situations have been found where noise plays a constructive role in a dynamical system. As in the prototypical example, all that is needed is a stochastic dynamics acting upon a system with two possible equilibrium states and a weak periodic forcing. In fact, those conditions can be further relaxed, an interesting extension being that of coherence resonance by which an excitable \cite{PK97} or chaotic \cite{PTMCG01} system can optimize the regularity of its firings under the correct amount of noise.  Similar results have been referred to as {\sl stochastic coherence} in ~\cite{ZGBU03} or {\sl stochastic resonance without external periodic force}~\cite{GDNH93,RS94}. Another extension considers the enhancement of the response to non-periodic forcings, mostly within the context of some biological applications \cite{collins}.

\begin{figure*}[t!]
\begin{center}
 \includegraphics[width=13cm,angle=0]{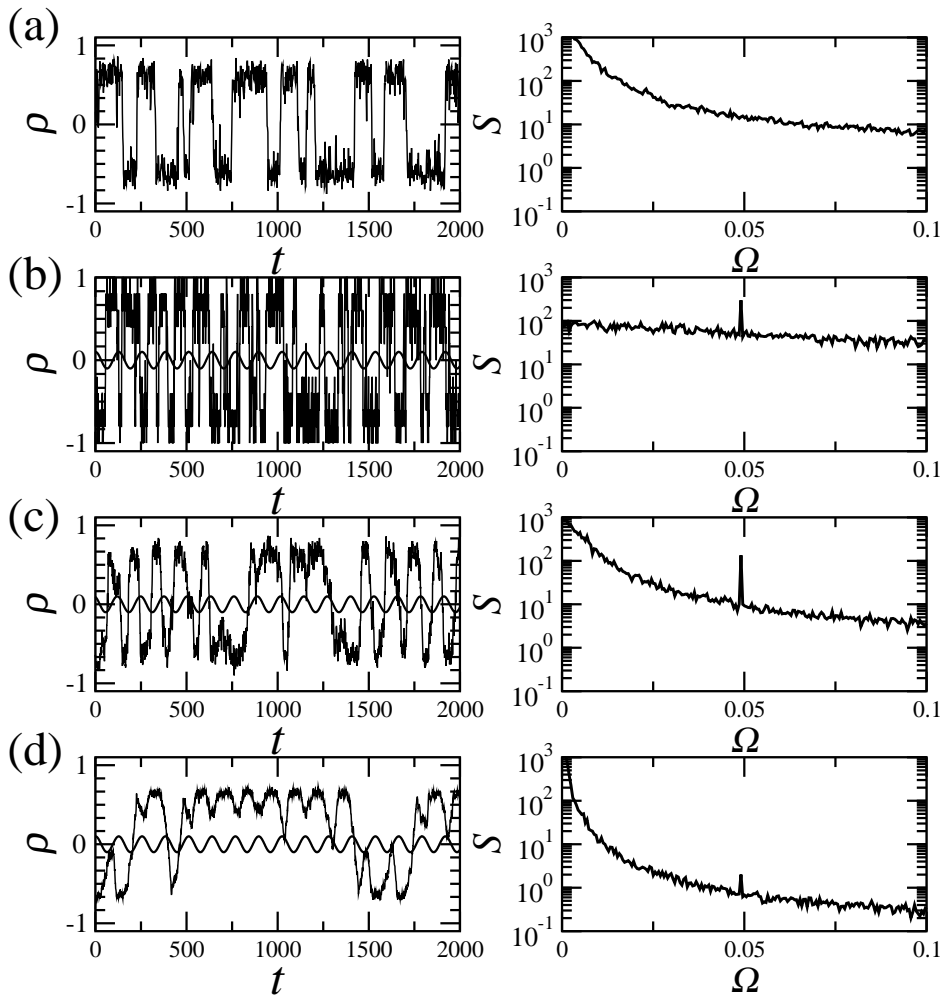}
\caption{\label{rhot} We plot the time-evolution of the average opinion
as a function of time (left column), and the power spectral density $S$
(right column). The first row (a) shows the dynamics in absence of an
external influence $\epsilon = 0$ in the case $N=100$. The last three rows, shows the
results for different system sizes: $N=10, 100, 1000$ (b, c, and d, respectively). The external signal
(the sinusoidal thin line in the evolution plots) has a period $T=128$. Note that this signal is better followed by
a system with an {\it intermediate} size (panel c). This is also signalled by the largest peak of the spectral density at the driving frequency: $\Omega = 2\pi / 128$.  
}
\end{center}
\end{figure*}

Concerning the dependence on system size relevant to this review,  Pikovsky and collaborators \cite{PZC:2002} showed that an extended system can achieve a resonance effect with respect to an external signal as a function of the size of the system in a phenomenon that has been named as {\sl system size stochastic resonance}. The explanation is that, due to the law of large numbers, the intensity of the noise experienced by the macroscopic variable scales as $N^{-1/2}$, being $N$ a measure of the system size. Therefore, the effective noise intensity for the, average, macroscopic variable can be controlled by varying the system size until the resonance condition between the Kramers time and the period of the forcing is satisfied. This very interesting result has been recently extended to the system size coherence resonance phenomenon, with a similar explanation \cite{TMG03}.

In the dynamics of social systems, stochastic resonance was found in very simple opinion models \cite{KZ:2002,WW:1991}. The basic ingredients are (i) the individuals of a society can have a binary opinion (in favor or against) a particular topic; this opinion can change by social interaction in such a way that both opinions are equally likely to prevail (this is the bistability condition); (ii) despite the interaction, individuals might decide to change their opinion in a random manner (stochastic terms in the dynamics); (iii) there is an external agent (advertising) that influences the opinion of the individuals (a forcing term). Given the aforementioned conditions, it should be of no surprise that a system size stochastic effect has been found in specific models. 

This result is quite general, and we now explain in detail a specific model which displays this effect. We have chosen a particularly simple opinion model developed by Kuperman and Zanette\cite{KZ:2002}. In this models each of the $i=1,\dots, N$ individuals of a society can have a binary opinion, say $\mu_i=\pm 1$, on a given topic. The individuals interact through a network of neighbors. We have considered both the Watts-Strogatz small world networks as wel as the scale-free network, constructed with the Barab\'asi-Albert algorithm \cite{Barabasi99a}.

The first ingredient of the dynamical evolution of the opinion is particularly simple and it is based upon a majority rule: at a given time step $t$ an individual $i$ is selected, this individual then adopts the majority opinion held by his set of neighbors (there is no bias rule: in case of a tie, a random value for $\mu_i$ is selected). This step, acting alone, would lead to a society holding a single opinion, either $+1$ or $-1$, i.e, the dynamics is bistable. The selected value, $+1$ or $-1$, depends on the initial condition and the particular set of individuals chosen at each time step. 

The second ingredient reflects the existence of an external influence agent such as advertising: with probability $|\epsilon\cos(\omega t)|$ the opinion of the randomly chosen individual $i$ changes to $\rm{sign}(\cos(\omega t))$. Here, $\omega=2\pi/T$ is the frequency with which the externally favored opinion changes with time. This periodic variation aims to reflect in some simplified manner the periodic variation of fashion.

The third and final ingredient of the dynamics is random choice. Whatever the results of the two previous steps, individual $i$ decides with a probability $\eta$ to change his opinion to a randomly chosen value.

\begin{figure}[t!]
\begin{center}
 \includegraphics[width=6.5cm,angle=-90]{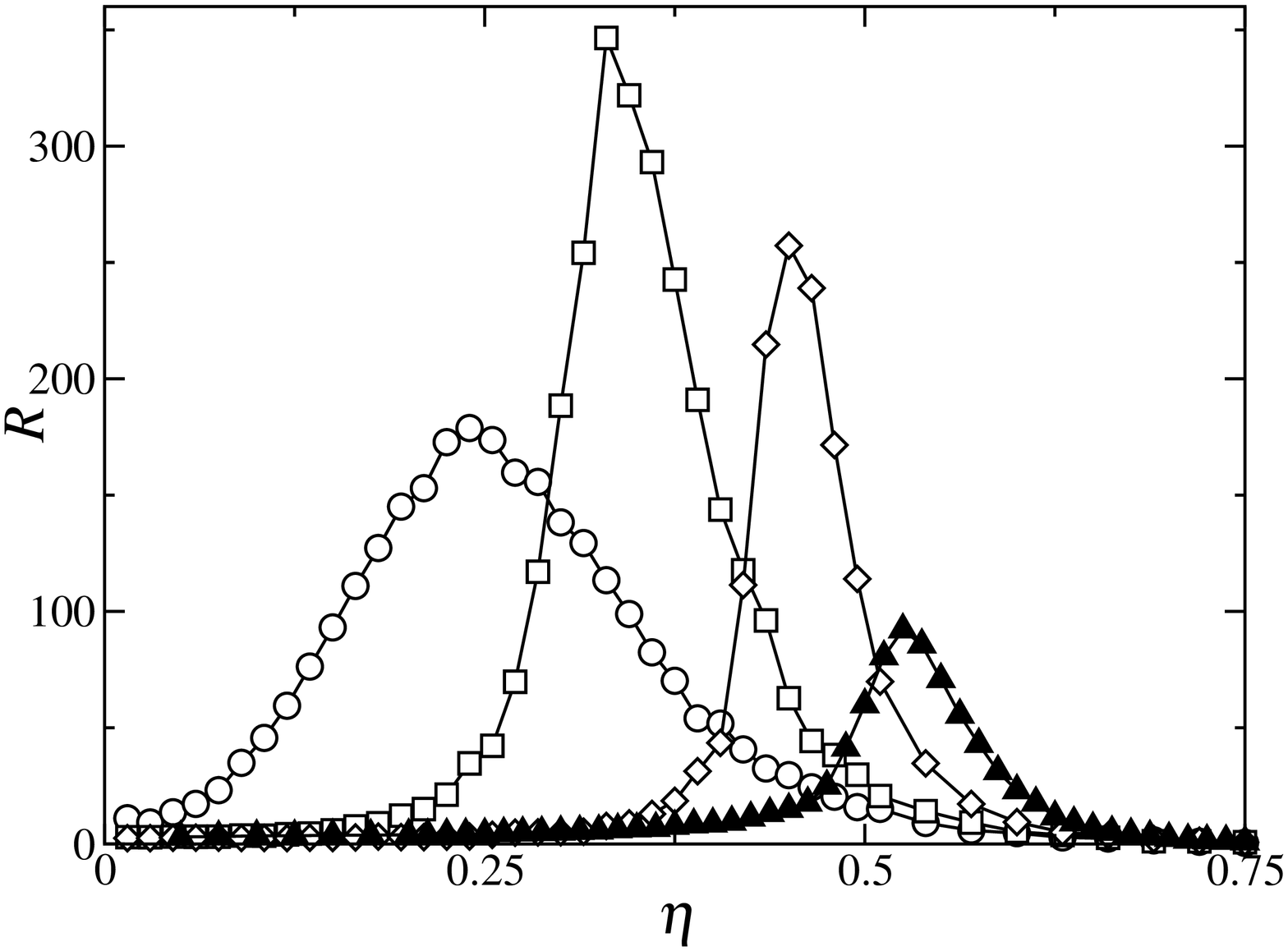}
\caption{\label{rhotn} We plot the system response (measures through the spectral amplification factor $R$) for different network topologies in the Kuperman-Zanette model for opinion spreading. The different curves correspond to different network topologies. The curves with open symbols correspond to a regular network  (circles), a small world networks with rewiring probability $p = 0.20$ (squares) and a fully random network, or $p = 1$ (diamonds). The black triangles correspond to a Bar\'abasi-Albert scale free network. The lines are only a guide for the eyes. The system parameters are $N = 2\times 10^3$, $T_s = 128$, $\epsilon = 0.02$.}
\end{center}
\end{figure}

After these three dynamical steps are taken consecutively (in the order specified above) by individual $i$, time increases by a time step $t\to t+1/N$.

We  look at the average opinion $\rho=N^{-1}\sum_i\mu_i$. In the absence of the periodic forcing, $\epsilon=0$, it is found that $\rho$ fluctuates stochastically between two values close to $+1$ and $-1$. In fact, the typical evolution, see the top panel of figure \ref{rhot} is very similar to the stochastic dynamics of a particle in a double well potential. A characteristic switching (Kramers) time can be defined as the average time to jump from one favored opinion to the other. It is found that this time decreases with noise intensity $\eta$ and increases exponentially with system size $N$. Although the dynamical rules do not allow us to identify a bistable potential function, it is clear that qualitatively the dynamics is such that a barrier between the two stable opinions has to be overcome in order for the society to change opinion. 

This analysis confirms that we have at hand in this simple opinion model all the ingredients necessary for stochastic resonance: a bistable stochastic dynamics with a typical switching time $T_K$. As this time decreases with noise rate, it is possible to find a value of $\eta$ such that half the external period coincides with the switching time, and hence the system responds optimally. This result was proven in reference \cite{KZ:2002}, although the first prediction for stochastic resonance in a model for opinion formation was obtained by Baninec in \cite{PB:1997}. We show this result in figure \ref{rhotn} where we plot  the spectral amplification factor, defined as  $R = 4 \epsilon^{-2} \left| \langle \hbox{e}^{i \, 2 \pi t/T} \rho(t) \rangle \right|^2$, ($\langle \dots\rangle$ denotes a time average). This is known to be a good measure of the response of the system to the forcing signal\cite{jung1}.

\begin{figure}[t!]
\begin{center}
 \includegraphics[width=6.5cm,angle=-90]{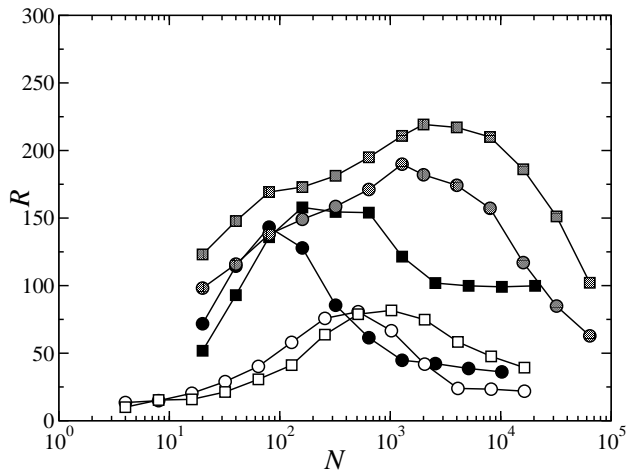}
\caption{\label{snrn} Spectral amplification factor as a function of $N$ for different network topologies: gray symbols show the results for a small-world network with rewiring probability $p=0.2$, black symbols are for a fully random network  and white symbols correspond to the Bar\'abasi-Albert network. In all the plots it is apparent an optimum response for an intermediate system size. The system parameters are as in the previous figures.}
\end{center}
\end{figure}

The fact that the switching time $T_K$ depends on the system size $N$, makes it possible to achieve the resonance condition by varying $N$. This interesting result shows that the influence that the forcing signal has on the average variable $\rho$ depends on the size $N$. In other words, that the effect of advertising is maximum for a population of the right size, neither too large nor too small. This result is shown in figure \ref{snrn} where it is shown that  the spectral amplification factor $R$ has a maximum as a function of the system size $N$. Although this result has been obtained within the framework of an specific model, it is important to realize that it wil also be present in other opinion models with the same basic ingredients. For instance, as shown in the figure \ref{snrn} the network structure is not a relevant variable here since the amplification factor always displays a maximum at an optimal value of the system size. Of course, the exact value of the optimal size does depend on the network structure.

\section{Conclusions}
\label{sec:conclusions}
In this paper, we have reviewed some situations in which system size can have a non-trivial role in some models of relevance to social systems. The first example was that of apparent phase transitions in the Galam's model of biased opinion. We showed in that example that there is a clear change of behavior in the sense that it is possible to identify a transition point separating the two possible final opinions. The transition point occurs at a {\sl pseudo-critical} value $p_c$ for the initial fraction of supporters of the favored opinion. However, the value of $p_c$ depends on the system size in such a way that $p_c$ tends to zero when the system size $N$ increases to infinity. Therefore, in a strict thermodynamic sense, the transition does no exists since it does not survive the thermodynamic limit. This is similar to what happens in one-dimensional models for the growth of rough surfaces, for which it can be proven that a change of behavior between a rough and a smooth phase occurs at some {\sl pseudo-critical} value of the temperature. However, in accordance with some general theorems, there can not be a true phase transition in this one-dimensional system. 

Next we have unveiled a similar apparent phase transition in Axelrod's model for culture formation. Here the apparent phase transition corresponds to a change of behavior in the existence of global or multicultural states as a function of the noise intensity, as measured by the existence of cultural drift. Again, although a clear change of behavior can be observed,  the transition does not exist in the true thermodynamic sense, since the {\sl pseudo-critical} value of the noise intensity in regular lattices scales as $r_c\propto 1/N\ln N$ in two dimensions or as $r_c\propto 1/N^2$ in one dimension. We have also considered the role of the structure of the social interactions by studying Axelrod's model in small world and scale free networks. It is found that in the Barab\'asi-Alert network, a pseudo-critical value $q_c$ for the heterogeneity parameter can be identified as separating the regimes of globalized and localized states. In the thermodynamic limit, $q_c$ tends to zero and the transition disappears.

Finally, we have presented a simple model for opinion formation based upon a majority rule in which it is shown that the effect that an external forcing (such as advertising) has on the society depends on the population size. A society of the proper size, neither too large nor too small, would react optimally to a weak signal.

This work has stressed the non-trivial role that the system size has in the dynamics of social systems. There are changes of behavior which appear only in finite systems. In a strict sense, these changes of behavior can not be called {\sl phase transitions} since they disappear in the thermodynamic limit. However, one can not forget that social systems are never in the thermodynamic limit and that those effects can not  be foreseen should one insist in taking this mathematical limit, as statistical physicists are routinely used to do.

{\bf Acknowledgements} This work has summarized the result of the direct interaction with P. Amengual, V. Eguiluz, K. Klemm, M.San Miguel and H. Wio, amongst others. We acknowledge financial support from MCyT (Spain) and FEDER (EU) through projects FIS2004-5073 and FIS2004-953.


\begin{thebibliography}{10}

\bibitem{pathria}
{R.K. Pathria}.
\newblock {\em Statistical Mechanics}.
\newblock Butterworth-Geinemann, 2nd edition, 1996.

\bibitem{cardy}
{J.L. Cardy}.
\newblock {\em Finite-{S}ize scaling}.
\newblock North-Holland, 1988.

\bibitem{weidlich}
{W. Weidlich}.
\newblock {\em Sociodynamics-{A} systematic approach to mathematical modeling
  in social sciences}.
\newblock Taylor \& Francis, London, 2002.

\bibitem{Ball}
{P. Ball}.
\newblock Utopia theory.
\newblock {\em Physics World}, October 2003.

\bibitem{stauffer04}
D.~Stauffer.
\newblock {\em Physica A}, 336:1, 2004.

\bibitem{schneider}
{T. Schneider and E. Stoll}.
\newblock {\em Phys. Rev. B}, 22:5317, 1980.

\bibitem{TS}
{T. Tsuzuki and K. Sasaki}.
\newblock {\em Prog. Theor. Phys. Suppl.}, 94:73, 1988.

\bibitem{Falo}
{F. Falo}, {A.R. Bishop}, {P.S. Lomdahl}, and {B. Horovitz}.
\newblock {\em Phys. Rev. B}, 43:8081, 1991.

\bibitem{sanbimo}
{A. S\'anchez}, {A.R. Bishop}, and {E. Moro}.
\newblock {\em Phys. Rev. E}, 62:3219, 2000.

\bibitem{CW2}
{S.T. Chui and J.D. Weeks}.
\newblock {\em Phys. Rev. B}, 14:4978, 1976.

\bibitem{knopf}
{H.J.F. Knops} and {L.W.J. den Oude}.
\newblock {\em Physica A}, 103:597, 1980.

\bibitem{nozi}
{P. Nozi\`eres}.
\newblock In {C. Godr\'eche}, editor, {\em Solids {F}ar from {E}quilibrium},
  Cambridge, 1991. Cambridge University.

\bibitem{acst03}
{S. Ares}, {J.A. Cuesta}, {A. S\'anchez}, and {R. Toral}.
\newblock {\em Phys. Rev. E}, 67:046108, 2003.

\bibitem{vanHove}
{L. van Hove}.
\newblock {\em Physica}, 16:137, 1950.

\bibitem{lieb}
{E.H. Lieb} and {D.C. Mattis}, editors.
\newblock {\em Mathematical {P}hysics in {O}ne {D}imension}, New York, 1966.
  Academic.

\bibitem{CS}
{J.A. Cuesta} and {A. S\'anchez}.
\newblock {\em J. Phys. A}, 35:2373, 2002.

\bibitem{ft93}
B.~Forrest and R.Toral.
\newblock {\em J. Stat. Phys.}, 70:703, 1993.

\bibitem{Gal03}
{S. Galam and B. Chopard and A. Masselot and M. Droz}.
\newblock {\em Eur. Phys. J. B}, 4:529, 2003.

\bibitem{Gal00}
{S. Galam and J.D. Zucker}.
\newblock {\em Physica A}, 287:644, 2000.

\bibitem{SG:2002}
S.~Galam.
\newblock {\em Eur. Phys. J. B}, 25:403, 2002.

\bibitem{SG:2003}
S.~Galam.
\newblock {\em Physica A}, 320:571, 2003.

\bibitem{complex1}
{G. Deffuant and D. Neau and F. Amblard and G. Weisbuch}.
\newblock {\em Adv. Complex Syst.}, 3:87, 2000.

\bibitem{complex1b}
{G. Weisbuch and G. Deffuant and F. Amblard and J.P. Nadal}.
\newblock {\em Complexity}, 7:55, 2002.

\bibitem{heg01}
{R. Hegselmann and U. Krausse}.
\newblock {\em J. of Artif. Soc. and Social Sim.}, 5:3, 2002.
\newblock http://www.soc.surrey.ac.uk/JASS/5/3/2.html.

\bibitem{sznajd01}
{K. Sznajd-Weron and J. Sznajd}.
\newblock {\em Int. J. Mod. Phys. C}, 11:1157, 2000.

\bibitem{sznajd01b}
{K. Sznajd-Weron}.
\newblock {\em Phys. Rev. E}, 66:046131, 2002.

\bibitem{sznajd01c}
{K. Sznajd-Weron and J. Sznajd}.
\newblock {\em Int. J. Mod. Phys. C}, 13:115, 2002.

\bibitem{sznajd02}
{F. Salnina and H. Lavicka}.
\newblock {\em Eur. Phys. J. B}, 35:279, 2003.

\bibitem{stauffer1}
{D. Stauffer and A.O. Souza and S. Moss de Oliveira}.
\newblock {\em Int. J. Mod. Phys. C}, 11:1239, 2000.

\bibitem{stauffer1b}
{D. Stauffer}.
\newblock {\em Int. J. Mod. Phys. C}, 13:315, 2002.

\bibitem{stauffer11}
{D. Stauffer}.
\newblock {\em Int. J. Mod. Phys. C}, 13:975, 2002.

\bibitem{stauffer2}
{D. Stauffer}.
\newblock {\em J. of Artificial Societies and Social Sim.}, 5:1, 2001.
\newblock http://www.soc.surrey.ac.uk/JASS/5/1/4.html.

\bibitem{stauffer2b}
{D. Stauffer}.
\newblock How to convince others?: {M}onte {C}arlo simulations of the {S}znajd
  model.
\newblock In {J.E. Gubernatis}, editor, {\em Proc. {C}onf. on the {M}onte
  {C}arlo {M}ethod in the {P}hysical {S}ciences}. AIP.
\newblock cond-mat/0307133.

\bibitem{stauffer2c}
{D. Stauffer}.
\newblock {\em Computing in Science and Engineering}, 5:71, 2003.

\bibitem{Castellano}
{C. Castellano}, {M. Marsili}, and {A. Vespignani}.
\newblock {\em Phys. Rev. Lett.}, 85:3536, 2000.

\bibitem{castellano2}
{D. Vilone}, {A. Vespignani}, and {C. Castellano}.
\newblock {\em Eur. Phys. J. B}, 30:399, 2002.

\bibitem{Klemm1}
{K. Klemm}, {V.M. Eguiluz}, {R. Toral}, and {M. San Miguel}.
\newblock {\em Phys. Rev. E}, 67:026120, 2003.

\bibitem{Klemm2}
{K. Klemm}, {V.M. Eguiluz}, {R. Toral}, and {M. San Miguel}.
\newblock {\em Phys. Rev. E}, 67:045101R, 2003.

\bibitem{Klemm3}
{K. Klemm}, {V.M. Eguiluz}, {R. Toral}, and {M. San Miguel}.
\newblock {\em J. Economic Dynamics and Control}, 29:321, 2005.

\bibitem{krapi01}
{P.L. Krapivsky} and {S. Redner}.
\newblock {\em Phys. Rev. Lett.}, 90:238701, 2003.

\bibitem{Mobilia1}
{M. Mobilia}.
\newblock {\em Phys. Rev. Lett.}, 91:028701, 2003.

\bibitem{Gal04b}
{S. Galam}.
\newblock {\em Physica A}, 333:453, 2004.

\bibitem{stauffer4}
{D. Stauffer} and {S.A. S\'a Martins}.
\newblock {\em Physica A}, 334:558, 2004.

\bibitem{stanley}
{H.E. Stanley}.
\newblock {\em Introduction to phase transitions and critical phenomena}.
\newblock Oxford U. Press, 1971.

\bibitem{TTAWS:2004}
C.J. Tessone, R.~Toral, P.~Amengual, H.S. Wio, and M.~San Miguel.
\newblock {\em Eur. Phys. J. B}, 39:535, 2004.

\bibitem{galam1}
{B. Chopard}, {M. Droz}, and {S. Galam}.
\newblock {\em Eur. Phys. J. B}, 16:Rapid Note, 575, 2000.

\bibitem{galam2}
{S. Galam}, {B. Chopard}, and {M. Droz}.
\newblock {\em Physica A}, 314:256, 2002.

\bibitem{Axelrod97}
{R. Axelrod}.
\newblock {\em J. Conflict Res.}, 41:203, 1997.

\bibitem{AxelrodBook}
{R. Axelrod}.
\newblock {\em The {C}omplexity of {C}ooperation}.
\newblock Princeton University Press, Princeton, 1997.

\bibitem{WS98}
D.J. Watts and S.H. Strogatz.
\newblock {\em Nature}, 393:440, 1998.

\bibitem{Barabasi99a}
{A.L.~Barab\'asi} and {R.~Albert}.
\newblock {\em Science}, 286:509, 1999.

\bibitem{victornet1}
{K. Klemm} and {V.M. Egu\'{\i}luz}.
\newblock {\em Phys. Rev. E}, 65:036123, 2002.

\bibitem{NN:1981}
C.~Nicolis and G.~Nicolis.
\newblock {\em Tellus}, 33:225, 1981.

\bibitem{BSV:1981}
R.~Benzi, A.~Sutera, and A.~Vulpiani.
\newblock {\em J. Phys. A}, 14:453, 1981.

\bibitem{JSP70}
{F. Moss}, {A. Bulsara}, and {M.F. Shlesinger}, editors.
\newblock {\em Stochastic Resonance in Physics, Biology}, volume~70 of {\em
  Proceedings of the NATO Advanced Research Workshop}. J. Stat. Phys., 1993.

\bibitem{1998_Gammaitoni_reviews}
{L. Gammaitoni}, {P. H\"anggi}, {P. Jung}, and {F. Marchesoni}.
\newblock Stochastic resonance.
\newblock {\em Rev. Mod. Phys.}, 70:223, 1998.

\bibitem{kramers}
{H.A. Kramers}.
\newblock {\em Physica (Utrecht)}, 7:284, 1940.

\bibitem{PK97}
A.~Pikovsky and J.~Kurths.
\newblock Coherence resonance in a noise-driven excitable system.
\newblock {\em Phys. Rev. Lett.}, 78:775, 1997.

\bibitem{PTMCG01}
C.~Palenzuela, R.~Toral, C.R. Mirasso, O.~Calvo, and J.D. Gunton.
\newblock Coherence resonance in chaotic systems.
\newblock {\em Europhys. Lett.}, 56(3):347--353, 2001.

\bibitem{ZGBU03}
A.~Zaikin, J.~Garc{\'\i}a-Ojalvo, R.~B\'ascones, E.~Ullner, and J.~Kurths.
\newblock Doubly stochastic coherence via noise-induced symmetry in bistable
  neural models.
\newblock {\em Phys. Rev. Lett.}, 90:030601, 2003.

\bibitem{GDNH93}
{Hu Gang}, T.~Ditzinger, C.Z. Ning, and H.~Haken.
\newblock Stochastic resonance without external periodic force.
\newblock {\em Phys. Rev. Lett.}, 71(6):807, 1993.

\bibitem{RS94}
{W. Rappel} and {S. Strogatz}.
\newblock {\em Phys. Rev. E}, 50:3249, 1994.

\bibitem{collins}
J.J. Collins, T.T. Imhoff, and P.~Grigg.
\newblock {\em J. Neurophysiol.}, 79:1879, 1998.

\bibitem{PZC:2002}
A.~Pikovsky, A.~Zaikin, and M.A. de~la Casa.
\newblock {\em Phys. Rev. Lett.}, 88:050601, 2002.

\bibitem{TMG03}
R.~Toral, C.~Mirasso, and J.~Gunton.
\newblock {\em Europhys. Lett.}, 61:162, 2003.

\bibitem{KZ:2002}
M.~Kuperman and D.H. Zanette.
\newblock {\em Eur. Phys. J. B}, 26:387, 2002.

\bibitem{WW:1991}
W.~Weidlich.
\newblock {\em Phys. Rep.}, 204:1, 1991.

\bibitem{PB:1997}
P.~Baninec.
\newblock {\em Phys. Lett. A}, 225:179, 1997.

\bibitem{jung1}
{P. Jung} and {P. H\"anggi}.
\newblock {\em Europhys. Lett.}, 8:505, 1989.


\end{thebibliography}
\end{document}